\documentclass{ifae}

\begin{document}

\begin{header}
\begin{flushright}
CERN-PH-TH-2004-174\\
\end{flushright}
  \title{Fragmentation of bottom quarks\\ in top quark decay}

  \begin{Authlist}
    G.~Corcella

\it{CERN, Department of Physics, Theory Division,
CH-1211 Geneva 23, Switzerland}
  
\end{Authlist}

  \begin{abstract}
  We review the main aspects of the 
fragmentation of bottom quarks in top quark decay.
The NLO b-quark energy  spectrum presents 
large mass logarithms $\ln(m_t^2/m_b^2)$, which can be resummed by the use of 
the approach of perturbative fragmentation functions. 
Large soft contributions in both coefficient function and initial condition of
the perturbative fragmentation function have been resummed as well.
Results on the energy distribution of b quarks and b-flavoured 
hadrons are finally presented in both $x$ and moment spaces.
  \end{abstract} 
  
\end{header}

\section{Introduction}
A reliable understanding of bottom quark fragmentation in top quark 
decay ($\rm{t}\to \rm{bW}$) will be fundamental to accurately measure the top 
properties, such as its mass $m_t$, 
at present and future high-energy colliders.
In fact, the uncertainty on bottom quark fragmentation is one of 
the sources of systematic error on $m_t$ at the Tevatron accelerator
\cite{tevatron}
and will play a crucial role in the reconstruction of $m_t$ from
final states with leptons and $\mathrm{J/\psi}$ at the LHC \cite{avto}.
 
In this paper we investigate bottom fragmentation in top decay within 
the framework of perturbative fragmentation functions \cite{mele}. 
We shall resum collinear logarithms $\sim\ln(m_t^2/m_b^2)$ and soft terms
that appear in the next-to-leading order (NLO) b-quark 
energy distribution. 
We shall present results on the b-quark energy spectrum in top quark decay
and investigate the impact of collinear and soft resummation.
Hadron-level results on b-flavoured hadrons will be shown
in $x_B$ and moment spaces.

\section{Collinear and soft resummation}
In Ref.~\cite{cm} NLO corrections
to top decay ${\rm{t}}(p_t)\to {\rm{b}}(p_b){\rm{W}}(p_W) 
({\rm{g}}(p_g))$ 
have been computed 
for a massive b quark, and the differential width $d\Gamma/dx_b$,
with $x_b$ being the normalized
b-quark energy fraction in the top rest frame, has been calculated.
The differential rate obtained in \cite{cm} exhibits large mass logarithms 
$\sim\alpha_S\ln(m_t^2/m_b^2)$ that need to be resummed in order to improve
the prediction.

Such contributions can be resummed by using the
perturbative fragmentation approach \cite{mele}, which, up to power 
corrections, factorizes the rate of
heavy-quark production into the convolution of a
coefficient function, describing the emission of a massless parton,
and a perturbative fragmentation function
$D(\mu_F,m)$, where $\mu_F$ is the factorization scale.
In the $\overline{\mathrm{MS}}$ factorization scheme we have:
\begin{eqnarray}
{1\over {\Gamma_0}} {{d\Gamma^b}\over{dx_b}} (x_b,m_t,m_b) &=&
\sum_i\int_{x_b}^1
{{{dz}\over z}\left[{1\over{\Gamma_0}}
{{d\hat\Gamma_i}\over {dz}}(z,m_t,\mu_F)
\right]^{\overline{\mathrm{MS}}}
D_i\left({x_b\over z},\mu_F,m_b \right)^{\overline{\mathrm{MS}}}} \nonumber \\
&&+ {\cal
O}\left((m_b/m_t)^p\right) \; ,
\label{pff}
\end{eqnarray}
In Eq.~(\ref{pff}) $\Gamma_0$ is the Born width of the process $\rm{t\to bW}$.
The ${\cal O}(\alpha_S)$ top decay coefficient function has been computed in
\cite{cm}.

The perturbative fragmentation function expresses the
transition of the massless parton into the massive quark, and its value
at any scale $\mu_F$ can be obtained by solving the
Dokshitzer--Gribov--Lipatov--Altarelli--Parisi (DGLAP) evolution
equations \cite{ap,dgl} once an initial condition at a scale
$\mu_{0F}$ is given.

In \cite{mele} the NLO expression for $D(\mu_{0F},m)$,
which was argued to be
process independent, was given. 
The process independence has been lately established in
a more general way in Ref.~\cite{cc}.

The initial condition of the perturbative fragmentation function 
reads \cite{mele}:
\begin{equation}
D_p(x_b,\mu_{0F},m_b)=\delta(1-x_b)+{{\alpha_S(\mu_0)C_F}\over{2\pi}}
\left[{{1+x_b^2}\over{1-x_b}}\left(\ln {{\mu_{0F}^2}\over{m_b^2}}-
2\ln (1-x_b)-1\right)\right]_+.
\label{dbb}
\end{equation}

As discussed in \cite{cm}, solving the DGLAP equations for the 
evolution $\mu_{0F}\to \mu_F$, with a NLO kernel,
allows one to resum leading logarithms (LL)  
$\sim\alpha_S^n\ln^n(\mu_F^2/\mu_{0F}^2)$ 
and next-to-leading logarithms (NLL) $\sim\alpha_S^n
\ln^{n-1}(\mu_F^2/\mu_{0F}^2)$ (collinear resummation).
If we set $\mu_{0F}\simeq m_b$ and $\mu_F\simeq m_t$, we resum
large logarithms $\sim \ln(m_t^2/m_b^2)$, which are indeed the terms 
appearing in the massive, unevolved, NLO
$d\Gamma/dx_b$.

Moreover, both the ${\overline{\mathrm{MS}}}$
coefficient function \cite{cm} and the initial condition
of the perturbative fragmentation function (\ref{dbb}) present, 
at ${\cal O}(\alpha_S)$, terms
that behave like $1/(1-x_b)_+$ or $[\ln(1-x_b)/(1-x_b)]_+$, which become large
for $x_b\to 1$, i.e. for soft-gluon radiation.
In Mellin moment space, such contributions correspond to behaviours
$\sim\ln N$ and $\sim\ln^2N$ respectively.

Soft contributions in the perturbative fragmentation function
are process-independent and have been resummed in \cite{cc} with 
NLL accuracy.
Soft terms
in the coefficient function are instead process-dependent.
Resummation of 
LL $\sim\alpha_S^n\ln^{n+1}N$ and NLL $\sim\alpha_S^n\ln^nN$
contributions to the top-decay coefficient function has been
performed in \cite{ccm} and we do not report here the formulae for the sake 
of brevity.

\section{Parton-level results}
We would like to present results on the b-quark energy distribution in
top decay and investigate the effect of collinear and soft resummation.
\begin{figure}[hbtp]
  \begin{center}
    \epsfig{file=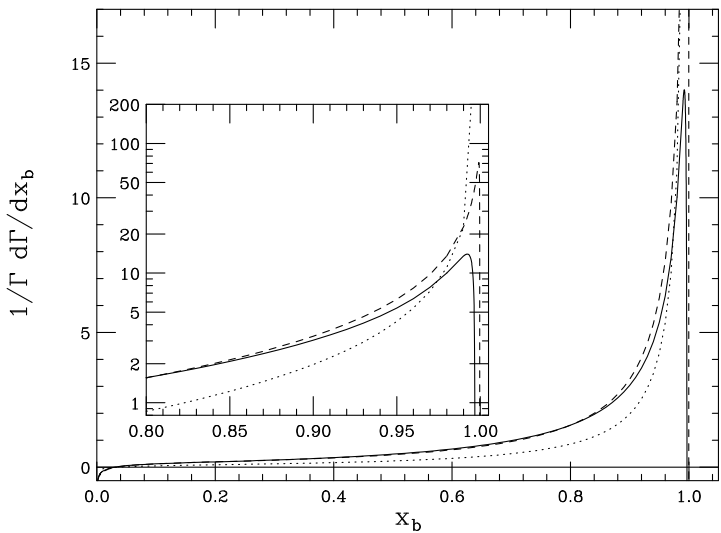,width=0.6\linewidth}
    \caption{b-quark energy distribution in top decay, according 
to the unresummed fixed-order calculation (dotted line), 
and after inclusion of collinear resummation
(dashed) and of both collinear and soft resummations (solid).
We have set $m_t=175$~GeV, $m_b=5$~GeV, $m_W=80$~GeV,
$\Lambda_{\mathrm{QCD}}=200$~MeV.
In the inset figure,  we show the same curves on a logarithmic scale,
for $x_b>0.8$.}
    \label{parton}
  \end{center}
\end{figure}
In Fig.~\ref{parton} we show the b-quark energy spectrum.
The NLO calculation 
lies below the two resummed predictions and 
is divergent as $x_b\to 1$. After the resummation of collinear terms 
$\sim\ln(m_t^2/m_b^2)$ the distribution exhibits a sharp peak at $x_b$ close to
1. Finally, the inclusion of soft-gluon resummation smoothens
out the distribution, which exhibits the so-called 
Sudakov peak. 

\begin{figure}[hbtp]
  \begin{center}
    \epsfig{file=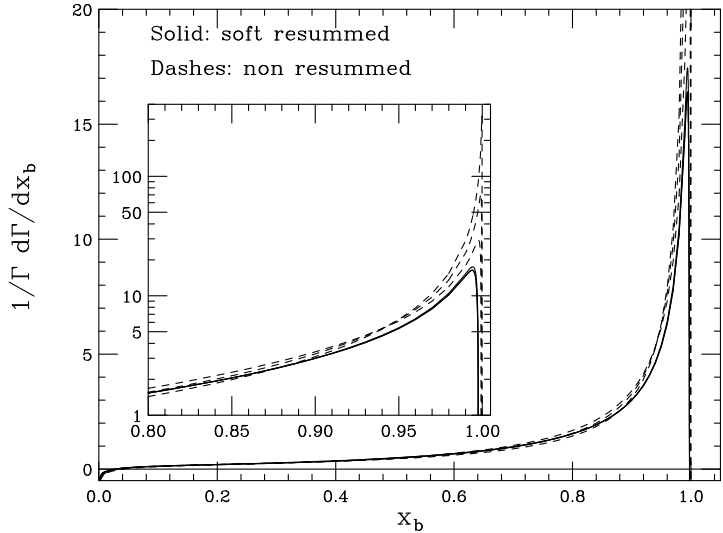,width=0.6\linewidth}
\caption{b-quark energy spectrum for different values
of the factorization scale $\mu_F$, with (solid) and
without (dashes)
NLL soft-gluon resummation.}
    \label{fac}
  \end{center}
\end{figure}

As discussed in \cite{ccm}, the implementation of collinear and
soft resummation
leads to a milder dependence of observables on the factorization 
and renormalization scales entering the calculation, which corresponds
to a reduction of the theoretical uncertainty.
As an example, Fig.~\ref{fac} shows the results on the dependence of the $x_b$
spectrum on the factorization scale $\mu_F$ in Eq.~(\ref{pff}), 
which is taken equal to 
$m_t/2$, $m_t$ and 2$m_t$, and the effect of soft resummation.
We note that while the unresummed prediction still exhibits a
dependence on the value chosen for $\mu_F$, the implementation
of soft resummation yields three almost undistinguishable distributions.

\section{Hadron-level results}

We would like to make predictions for the spectrum of b-flavoured hadrons 
in top decay. 
We write the normalized rate for the production of B-hadrons
B as a convolution
of the rate for the production of b quarks and
a non-perturbative fragmentation function $D^{np}(x)$:
\begin{equation}
{1\over {\Gamma}} {{d\Gamma^B}\over{dx_B}} (x_B,m_t,m_b)={1\over{\Gamma}}
\int_{x_B}^1 {{{dz}\over z}{{d\Gamma^b}\over {dz}}(z,m_t,m_b)
D^{np}\left({x_B\over z}\right)},
\label{npff}
\end{equation}
where $x_B$
is the B normalized energy fraction.
The parton-level rate $d\Gamma^b/ dz$ can be computed following
the method which has been discussed in the previous section.

As for the non-perturbative fragmentation 
function, one can use some phenomenological models with
tunable parameters, which are to be
fitted to experimental data.
We consider a power law with two parameters:
\begin{equation}
D_{np}(x;\alpha,\beta)={1\over{B(\beta +1,\alpha +1)}}(1-x)^\alpha x^\beta,
\label{ab}
\end{equation}
the model of Kartvelishvili et al. \cite{kart}
\begin{equation}
D_{np}(x;\delta)=(1+\delta)(2+\delta) (1-x) x^\delta
\label{kk}
\end{equation}           
and the non-perturbative fragmentation function of Peterson et 
al. \cite{peterson}:
\begin{equation}
D_{np}(x;\epsilon)={A\over {x[1-1/x-\epsilon/(1-x)]^2}}.
\label{peter}
\end{equation}
In Eq.~(\ref{ab}), $B(x,y)$ is the Euler beta function;
in (\ref{peter}) $A$ is a normalization constant.
We tune such models to $\rm{e}^+\rm{e}^-$ data from the 
ALEPH \cite{heister} and SLD \cite{abe} Collaboration. The ALEPH data 
refer to b-flavoured mesons, the SLD data to baryons and mesons.
When we do the fits, we must describe the $\rm{e}^+\rm{e}^-\to 
\rm{b}\bar{\rm{b}}$ process 
within the same framework as we did
for top decay, i.e. we use the perturbative fragmentation method,
NLL DGLAP evolution and NLL soft resummation.
As in \cite{cm,ccm}, we shall consider $x_B$ values within the range 
$0.18\leq x_B\leq 0.94$.
\begin{table}[ht!]
\caption{Results of fits of hadronization models to ALEPH and SLD data
on b-flavoured hadron production in $\rm{e}^+\rm{e}^-$ annihilation.}
\label{tab1}
\begin{center}
\begin{tabular}{||l|r|r||}\hline
&ALEPH\hspace{1.05cm} &SLD\hspace{1.45cm} \\ \hline 
\hspace{1.cm}$\alpha$&$0.51\pm 0.15$\hspace{0.79cm} &$2.04\pm 0.38$
\hspace{0.78cm} \\ \hline
\hspace{1.cm}$\beta$&$13.35\pm 1.46$\hspace{0.8cm} 
&$25.18\pm 3.27$\hspace{0.92cm} \\ \hline
$\chi^2(\alpha,\beta)$/dof&2.56/14\hspace{1.2cm} 
&11.50/16\hspace{1.2cm} \\ \hline
\hspace{1.cm}$\delta$&$17.76\pm 0.62$\hspace{0.8cm} 
&$16.59\pm 0.49$\hspace{0.92cm} \\ \hline
\hspace{0.2cm}$\chi^2(\delta)$/dof&10.54/15\hspace{1.2cm} 
&22.19/17\hspace{1.1cm}  \\ \hline
\hspace{1.cm}$\epsilon$&$(1.77\pm 0.16)\times 10^{-3}$
&$(1.61\pm 0.14)\times 10^{-3}$\\ \hline
\hspace{0.2cm}$\chi^2(\epsilon)$/dof&29.83/15\hspace{1.1cm}
&158.15/17\hspace{1.1cm} \\ \hline
\end{tabular}
\end{center}
\end{table}
\par In Table~\ref{tab1} we show the results of our fits, along with
the corresponding values of $\chi^2$ per degree of freedom. 
One can see that the power law with two
parameters (\ref{ab}) and the Kartvelishvili model (\ref{kk})
fit both ALEPH and SLD data rather well, while the Peterson fragmentation
function is marginally consistent with ALEPH and unable to
reproduce the SLD data.
Moreover, the values of
the best-fit parameters $\delta$ and $\epsilon$, fitted to ALEPH and SLD, are
in agreement within two standard deviations.

\begin{figure}[ht]
  \begin{center}
    \epsfig{file=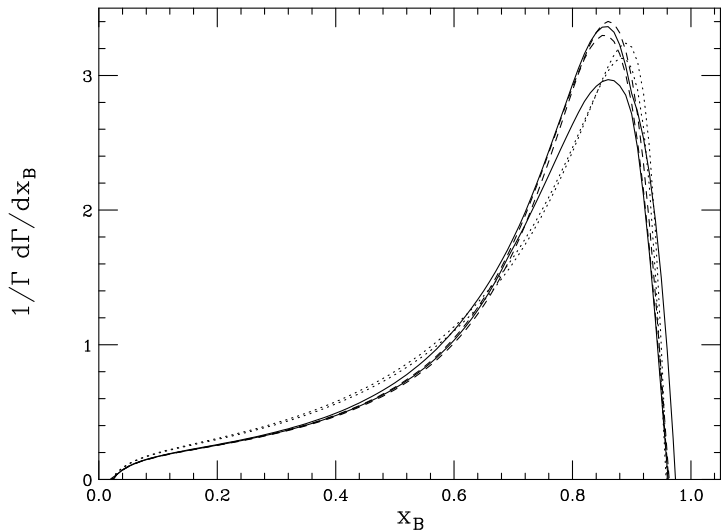,width=0.6\linewidth}
    \caption{B-hadron energy spectrum in top decay, according to the
power law (solid line), the Kartvelishvili (dashed) and
the Peterson model (dotted), fitted to the $\rm{e}^+\rm{e}^-\to 
\rm{b}\bar{\rm{b}}$
data from ALEPH. The plotted curves are the edges of bands at
one-standard-deviation confidence level.}
    \label{aleph}
  \end{center}
\end{figure}

In Fig.~\ref{aleph} we show our prediction for the B-hadron spectrum in
top decay, using all three hadronization models fitted to the ALEPH data.
In order to account for the uncertainties on the best-fit parameters,
for each model we plot a band corresponding to a
prediction at one-standard-deviation confidence level. 
From Fig.~\ref{aleph} we learn that the predictions based on the models
(\ref{ab}) and (\ref{kk}) are consistent, 
while the Peterson model yields a distribution that lies quite far from
the other two and is peaked at larger  values of $x_B$.

\begin{figure}[ht]
  \begin{center}
    \epsfig{file=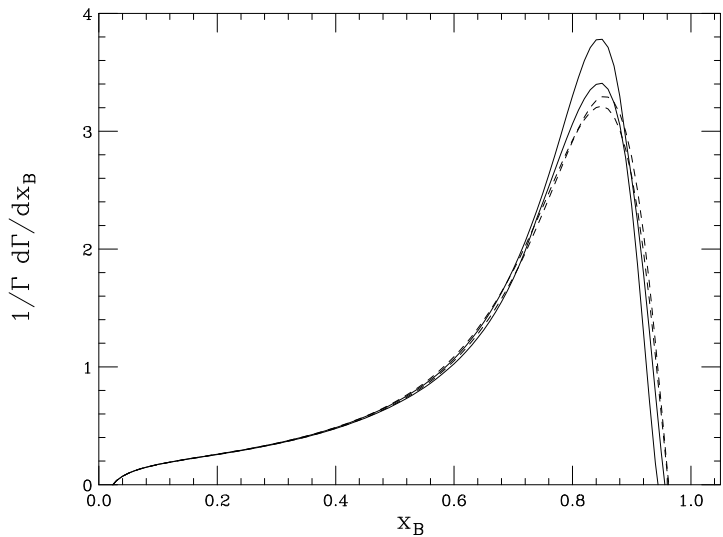,width=0.6\linewidth}
    \caption{As in Fig.~\ref{aleph}, but fitting the hadronization
models in Eqs.~(\ref{ab}) and (\ref{kk}) to the SLD data.}
    \label{sld}
  \end{center}
\end{figure}

In Fig.~\ref{sld} we plot the $x_B$ spectra yielded by models
(\ref{ab}) and (\ref{kk}), but fitted to SLD. 
Such distributions statistically agree at the confidence
level of two standard deviations.

\begin{figure}[ht]
  \begin{center}
    \epsfig{file=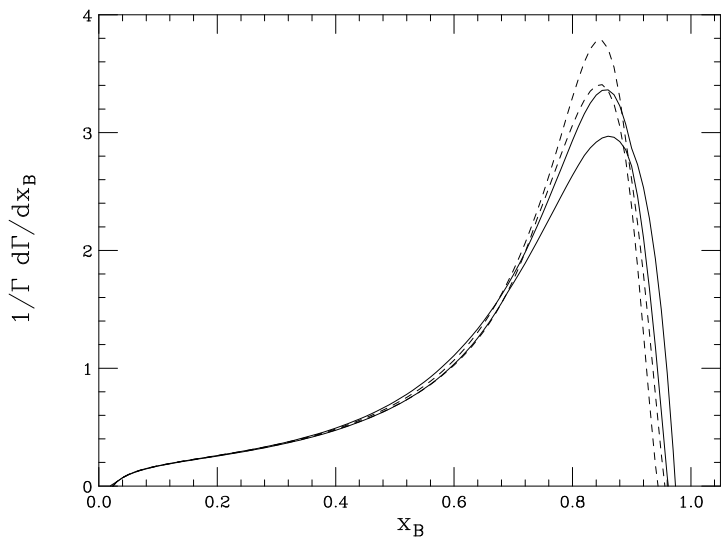,width=0.6\linewidth}
\caption{B spectrum in top decay according to the power law (\ref{ab}),
fitted to ALEPH (solid) and SLD (dashes) data.}
    \label{alsld}
  \end{center}
\end{figure}

In Fig.~\ref{alsld} we compare the predictions
obtained using the power law with two parameters, but fitted to 
ALEPH and SLD data. We observe that the spectra are distinguishable; this
difference may be related to the different hadron types that the
two experiments have reconstructed.
\par
We finally wish to present results on the moments of the B-hadron spectrum
in moment space $\Gamma_N^B$. Such moments 
can be written as the product of a 
perturbative and a non-perturbative contribution
$\Gamma^B_N = \Gamma^b_N D^{np}_N$. The advantage of working in moment space
is that one can extract $D^{np}_N$ from $\rm{e}^+\rm{e}^-$ data without relying
on any specific hadronization model.

Predictions for the moments $\Gamma^B_N$ of B-meson spectra in
top decay are given in Table~\ref{tab2}, where 
data from the DELPHI Collaboration \cite{delphi} are used to
obtain the non-perturbative information $D_N^{np}$.
Two sets of perturbative results ([A] and [B]) are shown,
the first using $\Lambda_{\mathrm{QCD}} = 0.226$~GeV and 
$m_b = 4.75$~GeV, the second
$\Lambda_{\mathrm{QCD}} = 0.2$~GeV and 
$m_b = 5$~GeV, the default values of this analysis. 
As expected, the
perturbative calculations and the corresponding non-perturbative
components differ at the level of few per cent, 
according to whether one uses set [A]
or [B]. However, the final hadron-level predictions for the physical results 
$\Gamma^B_N$ differ only  at the level of per mille.

\section{Conclusions}
We have considered bottom quark fragmentation in top quark decay
$\rm{t}\to \rm{bW}$.
We have pointed out that
the fixed-order result on the b-quark energy spectrum exhibits large
mass logarithms $\sim\ln(m_t^2/m_b^2)$, which can be resummed
to NLL accuracy using the approach of perturbative fragmentation
functions and DGLAP evolution equations. Moreover, 
NLL soft contributions to the coefficient function and
to the initial condition of the perturbative fragmentation function
have been resummed as well.

We have presented results on the b-quark spectrum in top decay, which
displays a remarkable impact of the inclusion of soft and
collinear resummation. In particular, the distributions exhibit
very little dependence on factorization and renormalization scales.

Predictions on b-flavoured hadron energy distributions in top decay have been
obtained using ALEPH and SLD data to parametrize some hadronization models
in $x_B$ space and DELPHI data to get non-perturbative information in
moment space.

The considered approach can now be applied to study several observables,
which are relevant to top quark phenomenology at the Tevatron and ultimately at
the LHC and compare the obtained results with the ones given by Monte Carlo
event generators.

\begin{table}[ht]
\caption{DELPHI data for the moments
$\sigma^B_N$, the resummed $\rm{e}^+\rm{e}^-$ perturbative
calculations for $\sigma^b_N$~\protect\cite{cc}, and
the extracted non-perturbative contribution
$D^{np}_N$. Using the perturbative results $\Gamma^b_N$, a prediction for
the physical observable moments $\Gamma^B_N$ is given.
Set [A]: $\Lambda_{\mathrm{QCD}} = 0.226$~GeV and $m_b = 4.75$~GeV,
set [B]: $\Lambda_{\mathrm{QCD}} = 0.2$~GeV and $m_b = 5$~GeV. }
\label{tab2}
\begin{tabular}{| c | c c c c |}
\hline
& $\langle x\rangle$ & $\langle x^2\rangle$ & $\langle x^3\rangle$
& $\langle x^4\rangle$ \\
\hline
$\rm{e}^+\rm{e}^-$ data $\sigma_N^B$&0.7153$\pm$0.0052 &0.5401$\pm$0.0064 &
0.4236$\pm$0.0065 &0.3406$\pm$0.0064  \\
\hline
\hline
$\rm{e}^+\rm{e}^-$ 
NLL $\sigma_N^b$ [A]   & 0.7666 & 0.6239 & 0.5246 & 0.4502  \\
$\rm{e}^+\rm{e}^-$ NLL $\sigma_N^b$ [B]   
& 0.7801 & 0.6436 & 0.5479 & 0.4755  \\
\hline
$D^{np}_N$ [A]          & 0.9331 & 0.8657 & 0.8075 & 0.7566 \\
$D^{np}_N$ [B]          & 0.9169 & 0.8392 & 0.7731 & 0.7163 \\
\hline
\hline
t-decay NLL $\Gamma^b_N$ [A]& 0.7750 & 0.6417 & 0.5498 & 0.4807 \\
t-decay NLL $\Gamma^b_N$ [B]& 0.7884 & 0.6617 & 0.5737 & 0.5072 \\
\hline
t-decay $\Gamma^B_N$ [A]        & 0.7231 & 0.5555 & 0.4440 & 0.3637 \\
t-decay $\Gamma^B_N$ [B]        & 0.7228 & 0.5553 & 0.4435 & 0.3633 \\
\hline
\end{tabular}
\end{table}

\section*{Acknowledgements}
The results here presented have been obtained in collaboration with
A.D. Mitov and M. Cacciari.

\end{document}